\documentclass[a4paper,prd,superscriptaddress,titlepage,linenumbers,aps,floatfix,notitlepage,twocolumn]{revtex4}
\usepackage[svgnames]{xcolor}
\usepackage[pdftex, unicode, colorlinks, bookmarks=true, citecolor=blue, link color=blue, urlcolor=blue]{hyperref}
\usepackage{amsmath, amssymb, amsfonts}

\usepackage{graphicx}
\graphicspath{{./images/}}

\newcommand{\bra}[1]{\langle#1|}
\newcommand{\ket}[1]{|#1\rangle}

\usepackage[normalem]{ulem}

\begin{document}
\title{A new type of quantum speed meter interferometer: measuring speed to search for intermediate mass black holes.}
\author{ Stefan L.~Danilishin}
\email{shtefan.danilishin@itp.uni-hannover.de}
\affiliation{Institut f\"ur Theoretische Physik, Leibniz Universit\"at Hannover and Max-Planck-Institut f\"ur Gravitationsphysik (Albert-Einstein-Institut), Callinstra\ss e 38, D-30167 Hannover, Germany}
\affiliation{SUPA, School of Physics and Astronomy, University of Glasgow, Glasgow G12 8QQ, United Kingdom}
\author{Eugene Knyazev}
\author{Nikita V. Voronchev}
\author{Farid Ya. Khalili}
\affiliation{M.V. Lomonosov Moscow State University, Faculty of Physics, Moscow 119991, Russia}

\author{Christian Gr\"af}
\affiliation{SUPA, School of Physics and Astronomy, University of Glasgow, Glasgow G12 8QQ, United Kingdom}
\author{Sebastian Steinlechner}
\affiliation{Institut f\"ur Laserphysik und Zentrum f\"ur Optische Quantentechnologien der Universit\"at Hamburg, Luruper Chaussee 149, 22761 Hamburg, Germany}
\author{Jan-Simon Hennig} 
\author{Stefan Hild}
\affiliation{SUPA, School of Physics and Astronomy, University of Glasgow, Glasgow G12 8QQ, United Kingdom}

\begin{abstract}
The recent discovery of gravitational waves (GW) by LIGO  has  impressively launched the novel  field of gravitational astronomy and it allowed us to glimpse at exciting objects we could so far only speculate about.  Further sensitivity improvements at the low frequency end of the detection band of future GW observatories rely on
quantum  non-demolition (QND) methods to suppress fundamental  quantum fluctuations of the light fields used to readout 
the GW signal.  Here we  invent a novel concept of how to turn a conventional Michelson interferometer into a QND speed meter interferometer with coherently suppressed quantum back-action noise by using  two orthogonal polarisations of light and  an optical circulator  to couple them. We carry out a detailed analysis of how imperfections and optical loss influence the achievable sensitivity and find that the configuration proposed here would significantly enhance the low frequency sensitivity and increase the observable event rate of binary black hole coalescences in the range of $10^2-10^3 M_\odot$  by a factor of up to $\sim300$.
\end{abstract}

\maketitle

\section{Main.}

The recently reported breakthrough espial of  gravitational waves emitted by coalescing binary black holes marked the starting point of the new field of gravitational wave astronomy \cite{GW_Discovery_Paper_PhysRevLett.116.061102}. The observations of Advanced LIGO gave evidence to a new population of black holes not consistent with our previous knowledge based on X-ray observations \cite{PhysRevX.6.041015_LVC}. Increasing the low frequency sensitivity of current and future gravitational wave observatories will not only allow us to improve the signal-to-noise ratio with which we can observe them, but also allow us to extend the observation capability to even heavier binary black hole systems. This would allow us to shed light onto many important questions, such as: What is the precise astrophysical production route of binary black hole  systems of tens of solar masses? What is the nature of spin-orbit and spin-spin couplings in coalescing binary black holes? Are the no-hair theorem and the second law of black hole mechanics valid?

In order to enhance the low frequency sensitivity of future gravitational wave detectors, a variety of noise sources has to be battled and improved, of which the most fundamental one is the so-called quantum noise, a inherent consequence of the quantum mechanics of the measurement process. 

Quantum fluctuations of the electromagnetic field were identified as the main fundamental limitation to the sensitivity of electromagnetic weak force sensors in the late 60s by Braginsky \cite{67a1eBr}. He showed that continuous monitoring of the test object position to infer an external weak force (\textit{e.g.}, GW) always leads to a quantum back-action of the meter on the probe object's position, thereby setting the \textit{standard quantum limit} (SQL) on the achievable precision of such a measurement. In interferometric sensors, like GW interferometers, light is used to monitor the distances between the mirrors.  Here back action noise originates from the quantum fluctuations of the light's intensity, leading to random radiation pressure  forces acting on the mirrors. The corresponding additional displacement noise is most pronounced at low frequencies due to the mirrors' dynamical response and stems from the fundamental quantum fluctuations of the light's phase, setting the imprecision of the position monitoring $\Delta x_{\rm imp}\propto 1/\sqrt{N_{\rm ph}}$ (here $N_{\rm ph}$ is the number of photons used for measurement) and the back action noise ($\Delta x_{\rm BA}\propto \sqrt{N_{\rm ph}}$).  Evidently, the na\"\i ve trade-off in $N_{\rm ph}$ yields the SQL that is the point where $\Delta x_{\rm imp} = \Delta x_{\rm BA}$. 

 The SQL stems from non-commutativity of the displacement as an operator at different times, \textit{i.e.} $[\hat x(t),\,\hat x(t')] \neq0$, which means that displacement measurement at a time $t$ will influence the result of the one at a later time $t'$. The observables that commute at different times and thus can be monitored continuously with arbitrary precision are known as \textit{quantum non-demolition (QND)} observables. The obvious choice for such observables are the conserved quantities of the test object, like energy, quadratures for the oscillator, or momentum for a free mass.

\begin{figure}
\includegraphics[width=.5\textwidth]{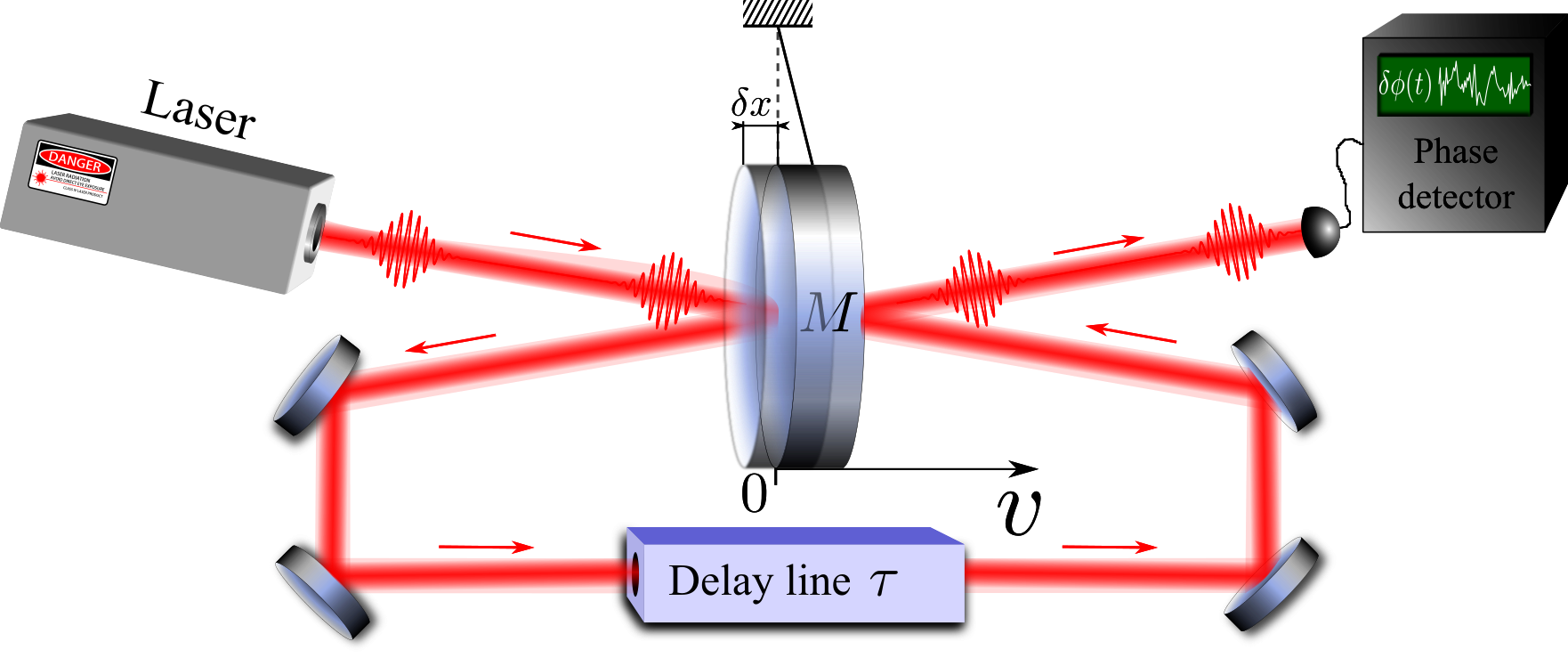}
\caption{ Conceptual scheme of optical speed measurement with two consecutive reflections of the light pulses from the front and the rear surfaces of the mirror.}\label{fig0}
\end{figure}

 Velocity measurement as a QND procedure proposed in \cite{1990_PLA.147.251_Braginsky_SM}, is based on the premise that at time-scales faster than the suspension pendulum period the mirror behaves as a free mass and its momentum is conserved and proportional to the velocity, $\hat p = m \hat v$. 
The more careful analysis has shown that the dynamics of the test object cannot be considered separately from that of the meter, which is the laser light in the case of GW interferometers. For a combined system `mirrors+light', the generalised momentum is rather a sum of two terms, $\hat P = m\hat v - g_{\rm SM}(t)\hat a_c$ than a simple proportionality to velocity (see, \textit{e.g.} Sec.~4.5.2 in \cite{Liv.Rv.Rel.15.2012}), where $g_{\rm SM}(t)$ is the strength of coupling between the light and the mirrors' mechanical motion, and $\hat a_c = (\hat a+\hat a^\dag)/\sqrt{2}$ is the amplitude quadrature of light (defined in terms of photon annihilation (creation) operator $\hat{a}$ ($\hat{a}^\dag$)).  Though sensing of the mirrors' velocity via outgoing light phase quadrature measurement is not a QND measurement, it nevertheless provides a substantial reduction of random back-action force \cite{Liv.Rv.Rel.15.2012}.

\begin{figure*}
\includegraphics[width=\textwidth]{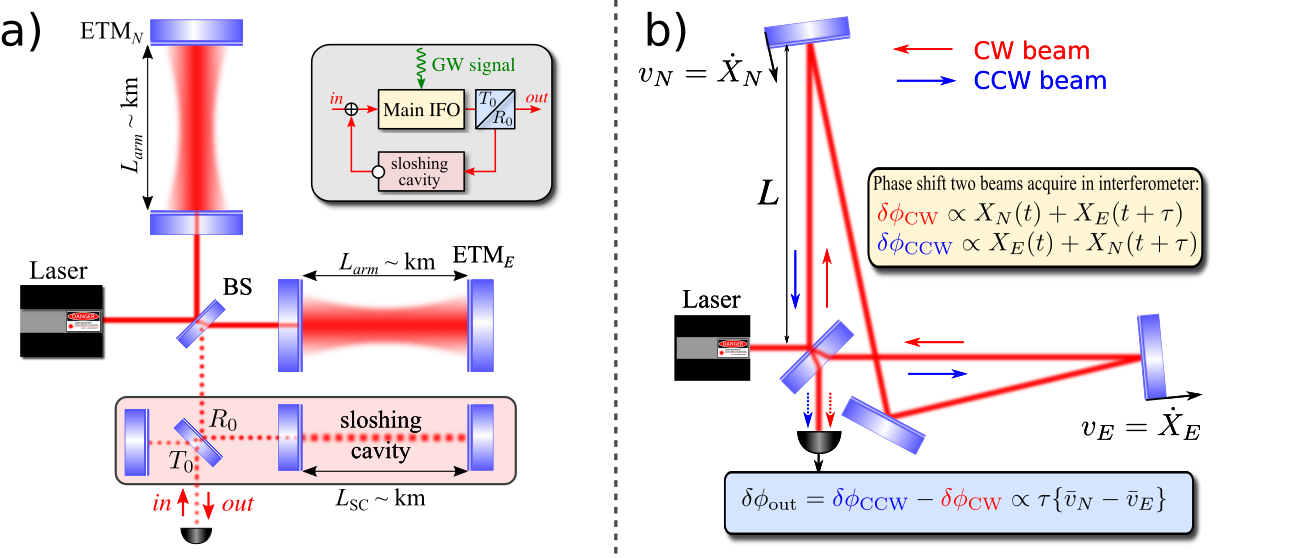}
\caption{ Two variants of implementation of speed meter interferometers, (a) the sloshing speed meter, and (b) the Sagnac speed meter. Inset in the grey rectangle in (a) represents the block diagram of the sloshing speed meter principle of operation. Here ETM stands for input test mass, BS is a beam splitter, and $T_0=1-R_0$ is the (power) transmissivity of the output coupling mirror.}\label{fig1}
\end{figure*}

 The simplest conceptual realisation of an optical speed meter is shown in Fig.~\ref{fig0} \cite{2004_Opt.Spec.96.727_Danilishin}. Here a laser sends short light pulses to the suspended mirror. The pulses, are reflected off the mirror twice with the time delay $\tau$ between the reflections. After each reflection the mirror's displacement is written in the phase of the pulse, hence after two reflections the pulse's phase is shifted by $\phi_{\rm pulse} \propto \hat x(t) - \hat x(t+\tau)\sim \tau\bar{v}$, where $\bar{v}$ is the mean velocity. 
Note that since the momentums transferred to the mirror by photons in the two reflections have opposite signs, and   since there is no decoherence between the reflections,  they compensate each other.  Therefore  quantum back-action noise  is suppressed  by $\sim\tau/T_{signal}\propto\Omega_{signal}\tau$, where $T_{signal} = 2\pi/\Omega_{signal}$ is the specific timescale of the signal force, \textit{e.g.}, the period of GWs.

This example grasps the two key features, which the  measurement scheme should possess to realise a speed measurement - (i) the probe (light) has to interact with the test object (mirror) twice, keeping coherence between the interactions (for coherent suppression of back-action noise), and (ii) the two terms in the interaction Hamiltonian that relate to the two consecutive measurements should have opposite signs.

 The first implementation for detection of gravitational waves was proposed in \textit{et al.} in \cite{00a1BrGoKhTh} (see Fig.~\ref{fig1}a) the traditional Fabry-P\'erot--Michelson was extended by an auxiliary ``sloshing'' optical cavity in the output port. This made the GW signal to "slosh" back and forth between the two coupled effective cavities with alternating phase. Hence, after the second pass through the interferometer, the outgoing light bears exactly the required combination of position signals, $\propto \hat x(t) - \hat x(t+\tau)\sim \tau\bar{v}$, yielding the speed measurement. This scheme was nick-named a "sloshing speed meter".
It has the distinctive feature that carrier and signal lights do not share the same optical path throughout the interaction, for the sloshing cavity is kept not pumped by a laser. This makes it very difficult to lock and control, and may also lead to signal loss from distortion in optical elements. A practical version of sloshing speed meter scheme was analysed in the great detail in \cite{Purdue2001,Purdue2002}.

Another solution was proposed by Chen in \cite{Chen2002}, demonstrating that a Sagnac interferometer with zero area performs a speed measurement.
Here the double measurement of the mirror position is performed naturally by two counter propagating light beams, which after the recombination on the beam splitter, produce the signal beam with phase dependent on the mean relative velocity of the end mirrors  (see Fig.~\ref{fig1}b and \cite{2004_Opt.Spec.96.727_Danilishin} for analysis). 

Quantum back-action noise suppression in both schemes owes to the fact that the radiation pressure force component which drives differential displacement of the arm mirrors, $x_{\rm dARM}=x_n-x_e$, stems from the beat note of the carrier classical amplitude $A\propto\sqrt{P_c}$ ($P_c$ is laser power circulating in the arms) with vacuum fields, $\hat{\pmb{i}}$, entering the readout port of the interferometer rather than with the laser fluctuations -- $\hat F^{\rm b.a.}(t)\propto A\,\hat i_c(t) $ ,  with $\hat i_c$ the amplitude quadrature of $\hat{\pmb{i}}$. In sloshing speed meters, the subtraction of two back-action kicks is provided by the $\pi$-phase shift that the dark port field  acquires after the reflection off the sloshing cavity, hence $\hat F^{\rm b.a.}\propto A\,\hat i_c(t) + e^{i\pi} A\,\hat i_c(t+\tau_{\rm sl}) = A\,(\hat i_c(t) - \hat i_c(t+\tau_{\rm sl}))$, and $\tau_{\rm sl}$ is the characteristic time of optical energy sloshing between the coupled cavities of the sloshing speed meter interferometer.

 In a Sagnac interferometer, the required "minus" sign  is provided by the $\pi$-phase difference between the reflected and the transmitted beam at the beam splitter. The suppression of quantum back-action here originates from the opposite sign of the radiation pressure forces from the clockwise and the counter clockwise propagating light beams, \textit{i.e.} $F_{\rm CW}^{\rm b.a.}+\hat F_{\rm CCW}^{\rm b.a.}\propto  A\,\hat{i}_c(t)-A\,\hat{i}_c(t+\tau_{\rm prop})$ with $\tau_{\rm prop}$ the light propagation time between the arms.

 The complexity of experimental implementation of these schemes led to the idea of using two orthogonal polarisations of light to separate the two beams sensing the mirrors in a Sagnac-type speed meter  \cite{2004_Phys.Rev.D.69.102003_Danilishin,PhysRevD.87.096008}. This approach allows to keep the km-scale arm cavities of the original Michelson unchanged, but it requires km-scale arm cavities of the original Michelson, but requires substantial modification to the input and output optics and the implementation of additional polarising elements of large physical dimensions, not used inside the core interferometers so far.

 An alternative scheme, proposed in \cite{PhysRevD.86.062001}, makes use of the differential optical mode of the Michelson interferometer with the polarisation orthogonal to that of the pumping laser as an effective sloshing cavity. The polarisation separation of the signal light fields from the "sloshing" ones is achieved by means of two quarter wave plates (QWP), 2 mirrors and a polarisation beam splitter (PBS). 

\begin{figure}[h]
\includegraphics[width=.45\textwidth]{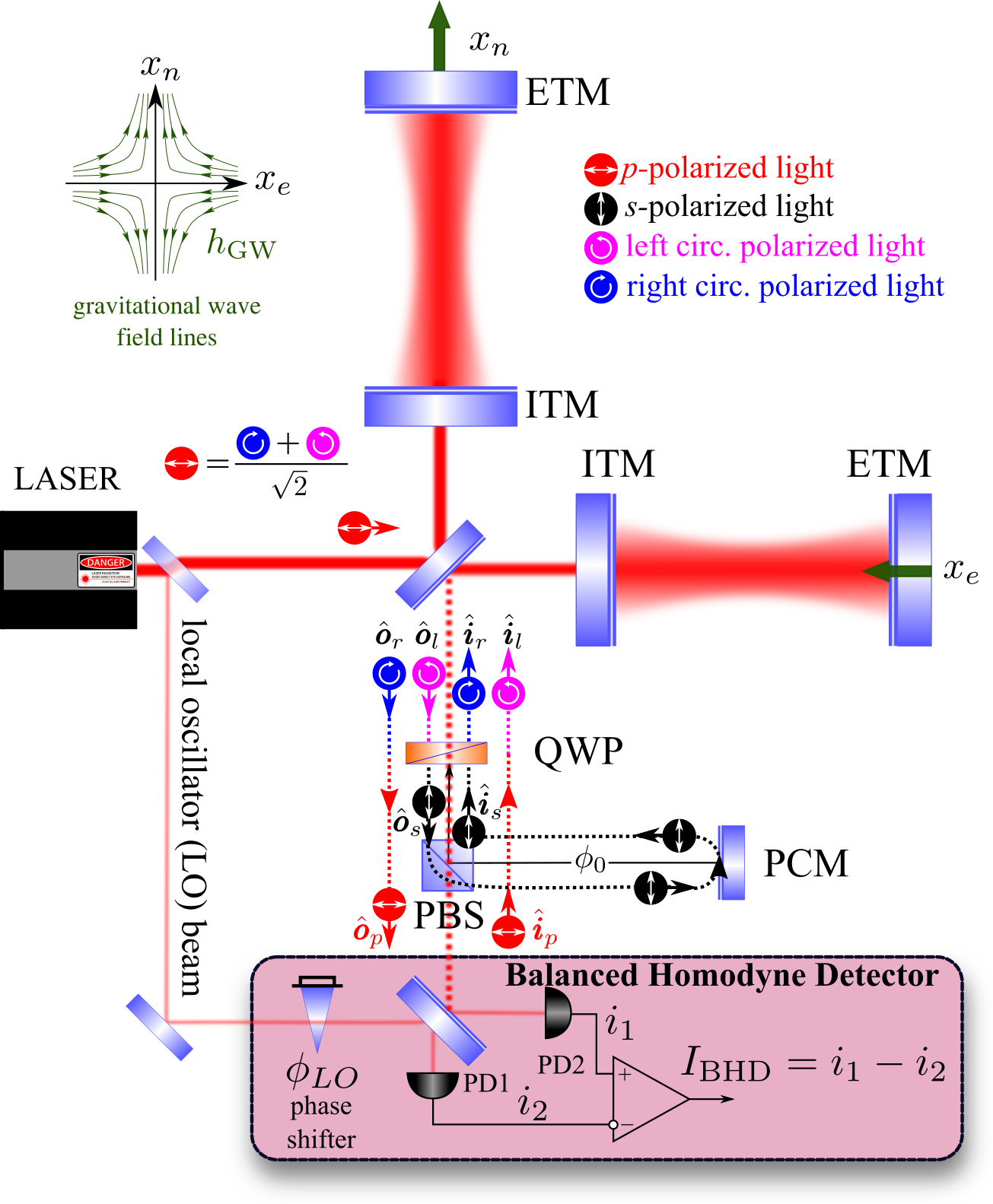}
\caption{Possible realisation of the polarisation circulation interferometer, using quarter-wave plate for polarisation separation. Here (E)ITM stands for (end) input test mass, PCM is a polarisation circulation mirror, QWP means quarter-wave plate, PBS stands for polarisation beam splitter and PD is a photodetector.}\label{fig2}
\end{figure}

\paragraph{Polarisation circulation interferometer as speed meter:} In this letter, we we propose a new, even simpler and hence more attractive scheme, where scheme where the two orthogonal polarisation modes of the Michelson interferometer serve as two counter propagating beams of a Sagnac-type interferometer. The scheme is shown in Fig.~\ref{fig2}. The main interferometer is pumped by the strong \textit{p}-polarised laser field $\pmb{p}_p$ that can be represented as a linear combination of two circularly polarised fields (marked by $(l)r$ for (counter)clockwise-polarised fields with polarisation basis vectors, $\vec{e}_{j}$ ($j=\{p,\,r\,,l\}$):
\begin{equation}
\pmb{p}_p\vec{e}_p = \pmb{p}_r\vec{e}_r + \pmb{p}_l\vec{e}_l\,,\ |\pmb{p}_r| = |\pmb{p}_l| = |\pmb{p}_p|/\sqrt{2}\,,
\end{equation}
 
 Coherent coupling between the two polarisations is performed by the \textit{polarisation circulator} comprising of  quarter-wave plate, PBS and the closing highly reflective mirror. PBS and QWP define the new circular polarisation basis for the light modes of the interferometer. The PBS lets through the \textit{p}-polarised vacuum field, $\hat{\pmb{i}}_p$, that is transformed by the QWP into the $l$-polarised field $\hat{\pmb{i}}_l$ . It enters the Michelson interferometer from the dark port and interacts optomechanically with the $\pmb{p}_l$ component of the pumping laser field $\pmb{p}$ and the differential mechanical degree of freedom of the interferometer mirrors, $x_{\rm dARM}(t) = x_n(t)-x_e(t)$. The outgoing $l$-polarised field $\hat{\pmb{o}}_l$, carrying information about the $x_{\rm dARM}$-displacement, is transformed into the \textit{s}-polarised field $\hat{\pmb{o}}_s$, which is reflected by the PBS towards the polarisation circulation mirror (PCM). The latter reflects $\hat{\pmb{o}}_s$ back towards the PBS where it arrives with an acquired phase shift $2\phi_0 = \pi$ and enters the main interferometer as $\hat{\pmb{i}}_r$ after being transformed by the QWP.  Delayed by the arm cavities ring-down time $\tau$, it senses the $x_{\rm dARM}(t+\tau) = x_n(t+\tau)-x_e(t+\tau)$, and couples with the $\pmb{p}_r$ component of the pumping laser field $\pmb{p}$.

 The $r$-polarised output field $\hat{\pmb{o}}_r$ leaves the readout port of the interferometer, transformed by the QWP into the \textit{p}-polarised one, $\hat{\pmb{o}}_p$ and transmitted by the PBS towards the balanced homodyne detector (BHD). The readout photocurrent is then proportional to the differential speed of the arms lengths change:
 \begin{equation}
 I_{\rm BHD}\propto x_{\rm dARM}(t+\tau) - x_{\rm dARM}(t) \simeq \tau\bar{v}_{\rm dARM}(t) \,.
 \end{equation}
  
 \begin{figure}[h]
\includegraphics[width=.5\textwidth]{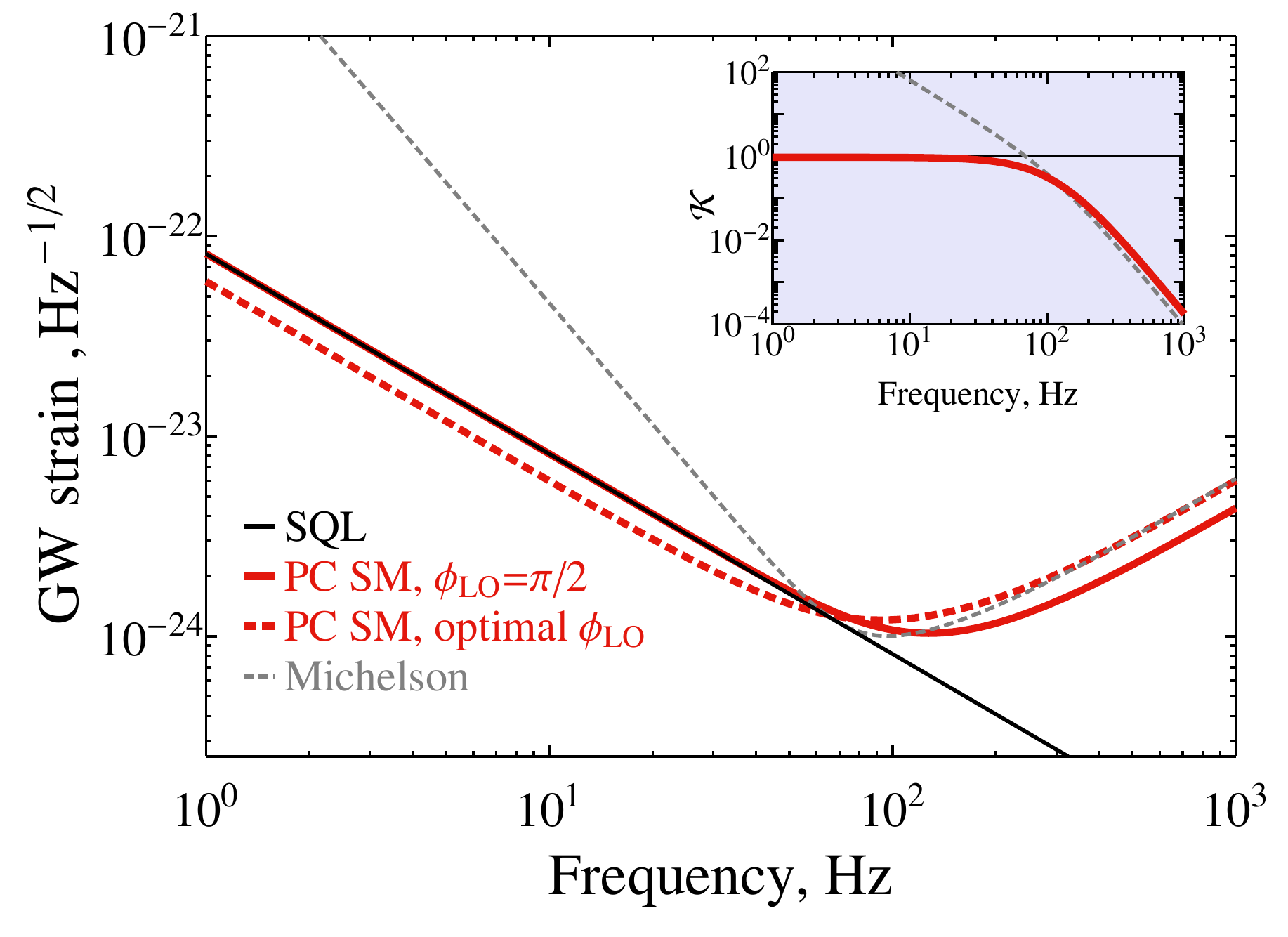}
\caption{ Quantum noise limited sensitivity (QNLS) of the polarisation circulation (PC) speed meter (\textit{red traces}) as compared to the QNLS of the equivalent Michelson interferometer (\textit{grey dashed trace}). \textit{Red dashed curve} shows the potential of speed meter to beat the free mass SQL (\textit{black trace}) in a wide band if optimal readout quadrature is measured: $\cot\phi_{\rm LO} = \mathcal{K}_{\rm PCSM}(0)$. In all other cases we assume \textit{phase readout}, \textit{i.e.} $\phi_{\rm LO} = \pi/2$. 
Hereafter we assume mirrors mass $M=200$~kg, power, circulating in each arm $P_c = 3$~MW, laser wavelength $\lambda_0 = 2\ \mathrm{\mu m}$ and effective interferometer bandwidth $\gamma/2\pi = 115$~Hz. .
}\label{fig:QNLS}
\end{figure}

\paragraph{Quantum noise limited sensitivity.} In order to give a more quantitative account of the quantum noise behaviour of the proposed scheme, we use the two-photon formalism of quantum optics \cite{85a1CaSch} (see Methods for details).  Here variations of two conjugate quadratures of the light field from the mean value are given by a 2D vector $\hat{\pmb{i}} \equiv \{\hat i_c,\,\hat i_s\}^{\rm T}$ of the \textit{amplitude} and the \textit{phase} quadrature operators, respectively.
Analysis of quantum noise of any interferometer starts from deriving the relations between the input and output light quadrature amplitudes, or \textit{I/O-relations} for sideband fields at an off-set frequency $\Omega = \omega - \omega_p$. For the lossless interferometer tuned in resonance, so as the GW signal shows up in the phase quadrature only, the general shape of I/O-relations is very simple \cite{02a1KiLeMaThVy}:
\begin{align}
\hat o_{p,c}(\Omega) &= e^{2i\beta}\hat i_{p,c}(\Omega)\,,\\
\hat o_{p,s}(\Omega) &= e^{2i\beta}\bigl[\hat i_{p,s}(\Omega)-\mathcal{K}\hat i_{p,c}(\Omega)\bigr]+e^{i\beta}\sqrt{2\mathcal{K}}\frac{h(\Omega)}{h_{\rm SQL}}\,,\label{eq:I/O_ph_quad}
\end{align}
where $\mathcal{K}(\Omega)$ is an optomechanical coupling factor describing the interaction of the mechanical degrees of freedom of the interferometer with light, $\beta(\Omega)$ is the frequency-dependent phase shift acquired by sideband fields as they pass through the interferometer, and $h_{\rm SQL} = \sqrt{\frac{8\hbar}{M L^2\Omega^2}}$ stands for the GW strain standard quantum limit for the effective mechanical mode of the interferometer with reduced mass $M$ and arm length $L$. The second term in the brackets in \eqref{eq:I/O_ph_quad} originates from the radiation pressure force driven by amplitude fluctuations. The last term in \eqref{eq:I/O_ph_quad} describes the response of the interferometer to the GW signal with strain $h(\Omega) = 2 x_{\rm dARM}(\Omega)/L$.

Using the matrix representation, outlined in the Methods, one can derive the quantum noise power spectral density (PSD) from the above I/O-relations that reads:
\begin{equation}\label{eq:Sh_tuned}
S^h(\Omega) = \dfrac{h^2_{\rm SQL}}{2}\Bigl\{\dfrac{1+[\mathcal{K}(\Omega)-\cot\phi_{\rm LO}]^2}{\mathcal{K}(\Omega)}\Bigr\}\,,
\end{equation}
where we assumed the homodyne readout of arbitrary quadrature defined by the local oscillator phase $\phi_{\rm LO}$.  

It is straightforward to derive the formulae for $\mathcal{K}$ and $\beta$ for any tuned configuration of interferometer. For a Michelson interferometer with total circulating power in each arm $P_c$, laser frequency $\omega_p = 2\pi c/\lambda_p$ and effective half-bandwidth $\gamma$ it reads $\mathcal{K}_{\rm MI}= \frac{2\Theta\gamma}{\Omega^2(\gamma^2+\Omega^2)}$ with $\Theta \equiv \frac{8\omega_pP_c}{McL}$, and frequency-dependent sidebands phase shift $\beta_{\rm MI} = \mathrm{arctan}\frac{\Omega}{\gamma}$. As shown in the Methods, the same expressions for the polarisation circulation interferometer in the speed meter regime reads:
\begin{equation}
\mathcal{K}_{\rm PCSM} = 2\mathcal{K}_{\rm MI}\sin^2\beta_{\rm MI} = \dfrac{4\Theta\gamma}{(\gamma^2+\Omega^2)^2}\,.
\end{equation}

The behaviour of $\mathcal{K}$ as function of frequency reflects the strength of interaction of light at this particular sideband frequency $\Omega$ with the mirrors of the interferometer. This includes both, the strength of back-action and the level of response one can expect from the particular scheme at a given signal frequency, as reflected by two terms in \eqref{eq:I/O_ph_quad} that contain $\mathcal{K}$. The inset to Fig.~\ref{fig:QNLS} shows clearly the differences between the Michelson and the PC speed meter in this regard. The sharp rise ($\propto\Omega^{-2}$) of $\mathcal{K}_{\rm MI}$ (grey trace) at low frequencies within the interferometer bandwidth, $\Omega<\gamma$, is responsible for worse quantum noise performance of the Michelson interferometer compared to the PC speed meter, characterised by the flat behaviour of $\mathcal{K}_{\rm PCSM}$ in that frequency region. This trend is responsible for the much improved speed meter quantum noise at low frequencies.  Moreover, as $\mathcal{K}_{\rm PCSM}(\Omega\to 0) = const$, one can improve low frequency sensitivity of the speed meter even more by choosing to measure the optimal quadrature by tuning the homodyne angle to $\phi_{\rm LO} =\mathrm{arc cot}\mathcal{K}_{\rm PCSM}(\Omega\to 0)$ as shown by the red dashed trace in Fig.~\ref{fig:QNLS}.

In a simple special case of $\phi_{\rm LO} = \pi/2$ the QNLS PSD is $S^h(\Omega) = h^2_{\rm SQL}\Bigl\{\mathcal{K}+\mathcal{K}^{-1}\Bigr\}/2$, and one can clearly see that $\mathcal{K}(\Omega_q) = 1$ is the condition of reaching the SQL. It defines the frequency $\Omega_q$, where QNLS curve touches the SQL, and therefore back-action and shot noise have equal contributions to the QNLS. For the Michelson interferometer, there is always a real solution to this condition, whereas for speed meter there is a threshold value of the ratio $\Theta/\gamma^3\geq1/4$ that sets the limit on the required circulating power for given interferometer bandwidth and \textit{vice versa}. For given half-bandwidth $\gamma$, the circulating power, required for the PC speed meter to reach the SQL is $P_c \geq McL\gamma^3/(16\omega_p)$.

\paragraph{Losses and imperfections analysis.}
\begin{figure}[h]
\includegraphics[width=.45\textwidth]{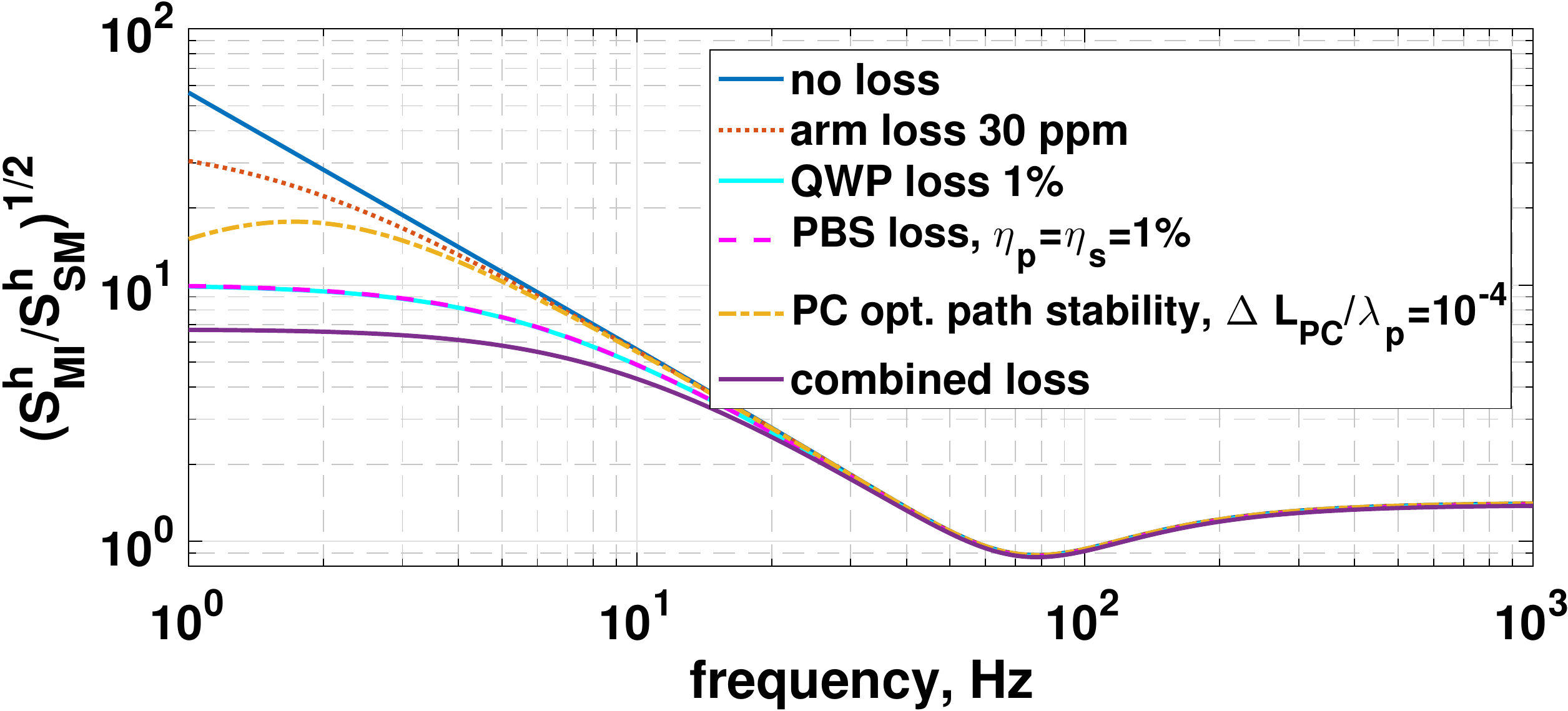}
\caption{ Influence of different sources of loss and imperfection on quantum noise-limited sensitivity of the polarisation circulation speed meter as compared to the equivalent Michelson interferometer.}\label{fig:loss}
\end{figure}

\begin{figure}[h]
\includegraphics[width=.45\textwidth]{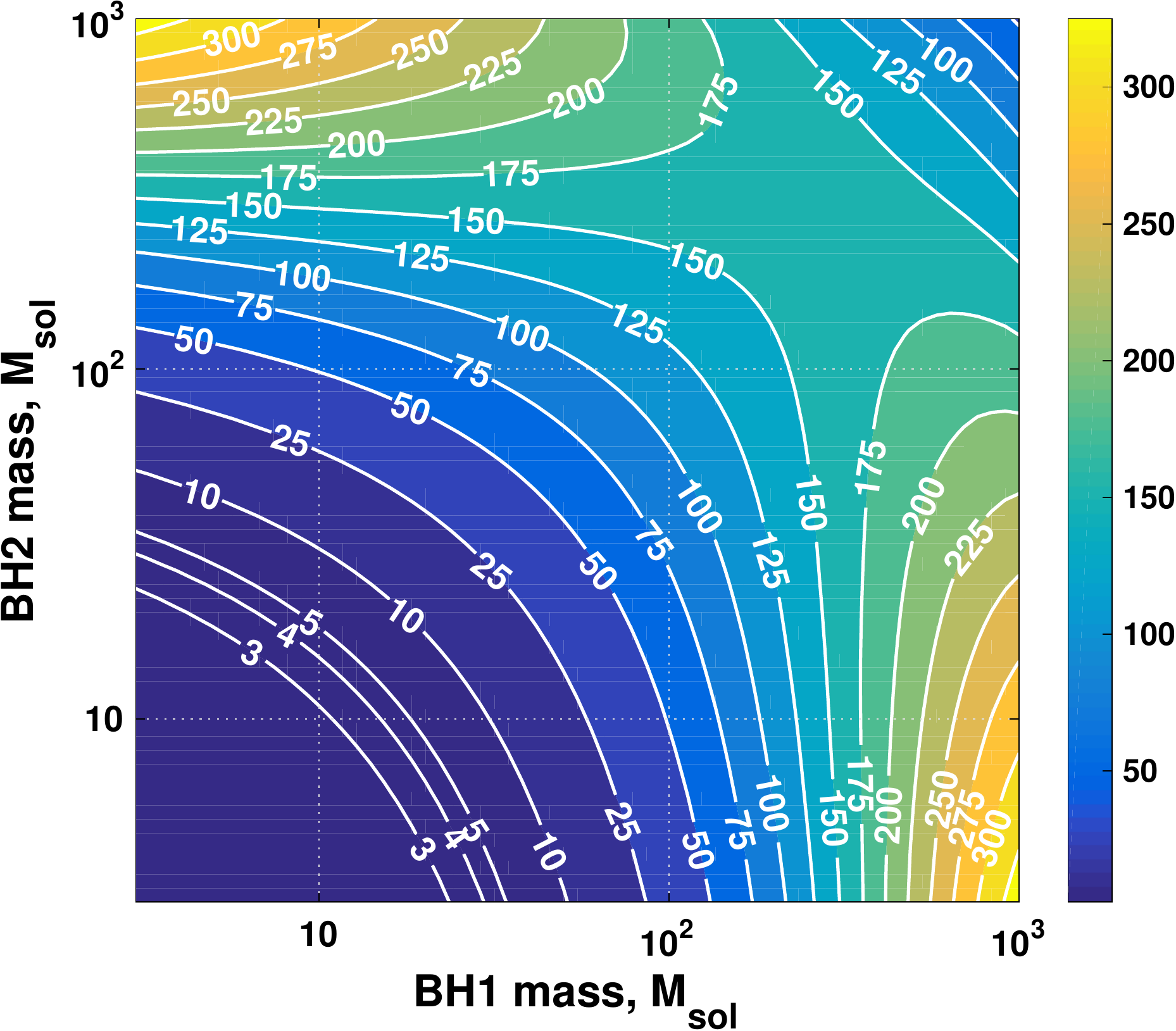}
\caption{Improvement in the anticipated  rate of detection of binary black holes (BBH) coalescenses (event rate) based on the  amplitude spectral density of the quantum noise of the proposed scheme as compared to the amplitude spectral density of the quantum noise of the equivalent Michelson interferometer.}
\label{fig:eventrateIF}
\end{figure}

To estimate the astrophysical potential of proposed scheme fairly, we need to assess the influence of the main sources of loss and imperfections of the real interferometer. 
In Fig~\ref{fig:loss}, we show the relative contributions (normalised by the QNLS of the equivalent lossless Michelson interferometer) losses and imperfections make to the realistic QNLS.

The leading source of loss for the proposed scheme is photon absorption and loss in the polarisation components, \textit{i.e.} absorption in the QWP (assumed single pass photon loss of $\epsilon_{\rm QWP} = 1\%$) and loss due to imperfect polarisation separation in the PBS (assumed extinction ratio for transmitted \textit{s}-polarised and reflected \textit{p}-polarised light of $\eta_s = \eta_p = 1\%$). One sees that both mechanisms contribute equally to the QNLS, which is no surprise as the input fields, $\hat{\pmb{i}}_p$, pass both elements the same number of times (4) before being read out at the output as $\hat{\pmb{o}}_p$. We also consider loss in the arm cavities, $\epsilon_{\rm arm}=30$~ppm as a realistic projection for the next generation GW interferometers. The arm loss influence at low frequencies, as shown by Kimble \textit{et al.}\cite{02a1KiLeMaThVy}, amounts to additional incoherent back action noise created by loss-associated vacuum fields. 

Finally, we analyse how robust the scheme is to the small deviations, $\Delta L_{\rm PC}\ll\lambda_p$, of the optical path length between the QWP and the PCM, defining $\phi_0$. Departure of $\phi_0$ from the $\pi/2$ value results in a partial leakage of the back-action term $\propto \mathcal{K}_{\rm MI}\hat i_{p,c}$ from the sine quadrature into the cosine one, thereby creating an additional back-action term in the quantum noise PSD $\propto2\mathcal{K}_{\rm MI}\Delta L_{\rm PC}/\lambda_p\propto 1/\Omega^4$ that leads, in conjunction with speed meter-like response ($\propto\Omega$), to a steep rise of noise at low frequencies, $\sqrt{S^h}\propto 1/\Omega^3$. This explains the downward bending of the corresponding yellow dash-dotted curve in Fig~\ref{fig:loss} (see Eq.~\eqref{eq:PCSM_QNLS_epsilon} in Methods).

\paragraph{Astrophysics results and prospects:}
A quantitative comparison of the QNLS of our proposed speed meter scheme and QNLS of an equivalent Michelson interferometer is shown in Fig~\ref{fig:eventrateIF}. (We assume for our analysis that due to the application of enhanced
techniques all other noise sources, such as Newtonian noise \cite{PhysRevD.30.732_Saulson}, seismic noise \cite{2015_CQG_32r5003M} and suspension thermal noise \cite{2012CQGra..29c5003C}, are pushed below the level of the QNLS).
For this we consider the realistic speed meter including the optical losses shown in Fig~\ref{fig:loss} and calculate the corresponding inspiral range (integrated for frequencies between 1Hz and the last stable orbit), i.e. the distance up to which we can observe the BH coalescence before its signal to noise ratio decreases below 8. Then we compare the speed meter inspiral range to the insprial range of an equivalent Michelson interferometer and derive the plotted improvement factor in terms of event rate, assuming a homogeneous sources distribution throughout the Universe.
We find for instance that for initial black hole masses similar to GW150914 \cite{GW_Discovery_Paper_PhysRevLett.116.061102} the speed meter would provide and improvement in the event rate of about 27. The largest improvement factors, however, are given for initial black holes in the range from $100 - 1000 M_\odot$ solar masses with obtained improvement factors larger than 100, which will open up access to investigating the potential existence of any intermediate mass BH population in this, so far unobserved, mass range.

\paragraph{Summary.}
In this article we suggested a new configuration for realising a quantum speed meter in laser-interferometric GW observatories.
 The key advantage of our configuration compared to other speed meter implementations, is that no additional optical components need to be implemented inside the main.
  The few additional components required to convert a standard advanced GW detector into our polarisation circulation speed meter can be placed in the output port of the interferometer (i.e. behind the signal extraction cavity).
   Our analysis showed that compared with a standard Fabry Perot Michelson interferometer our speed meter configuration provides significantly improved sensitivity at low frequencies. 
   Further, a detailed investigation was carried out to identify the influence of imperfections on the sensitivity. We found that the most critical factor is the optical loss of the quarter wave plate and the PBS.
    Using realistic values for imperfections and loss we found that the speed meter QNLS sensitivity would yield an improvement factor of larger than 100 in event rate for binary black hole mergers in the range from $10^2 - 10^3 M_\odot$. 
    Future analyses will focus on investigating further sensitivity improvements from adding additional complementary quantum noise reduction techniques, such as the injection of squeezed light states.  

\paragraph{Acknowledgements.} This work would not be possible without the support of and the insightful conversations with our colleagues from the Interferometry group of the Institute for Gravitational Research of the University of Glasgow. The authors are very grateful as well to our colleagues from the LIGO-Virgo Scientific Collaboration (LVC) for illuminating discussions and invaluable feedback on the research presented in this paper. S.H., S.L.D., C.G., J.-S.H, and S.S. were supported by the European Research Council (ERC- 2012-StG: 307245).  S.L.D. was supported by the Lower Saxonian Ministry of Science and Culture within the frame of “Research Line” (Forschungslinie) QUANOMET – Quantum- and Nano-Metrology. S.L.D. and S.H. were supported by the Royal Society International Exchanges Scheme Grant IE160125.  The work of E.K. and F.Y.K. was supported by the Russian Foundation for Basic Research Grants 14-02-00399 and 16-52-10069. F.Y.K. was also supported by the LIGO NSF Grant PHY-1305863. S.S. was supported by the European Commission Horizon 2020 Marie-Sk\l odowska-Curie IF Actions, grant agreement 658366.

\section{Methods}
\paragraph{Two-photon formalism:}
We calculate quantum noise of the proposed scheme using the so called two-photon formalism \cite{85a1CaSch}. This formalism is best suited for steady state analysis of quantum fluctuations in the optomechanical devices. In this formalism, the electric field strain of the plane electromagnetic wave of the laser beam with frequency $\omega_p$, cross-section area $\mathcal{A}$ and power $P_{\rm in}$ can be written as: $ \hat{E}(t) = \mathcal{E}_0\left[(A+\hat{a}_c(t))\cos\omega_p t+\hat{a}_s(t)\sin\omega_pt\right]$, where $\mathcal{E}_0 = \sqrt{4\pi\hbar\omega_p/(\mathcal{A}c)}$ is the second quantisation normalising constant, $A=\sqrt{2P_{\rm in}/(\hbar\omega_p)}$ is the carrier dimensionless amplitude, and $\hat a_{c,s}(t)$ stand for \textit{cosine} ("c") and \textit{sine} ("s") quadrature amplitudes of the quantum fluctuations with zero mean.

\paragraph{Input-output relations of the interferometer:}
In steady state, it is more convenient to describe quantum fluctuations in frequency domain by introducing a signal sideband frequency $\Omega\to\omega-\omega_p$ centred around the laser frequency $\omega_p$: $\hat a_{c,s}(t)=\int_{-\infty}^\infty \hat a_{c,s}(\Omega)e^{-i\Omega t}\,\frac{d\Omega}{2\pi}$. In this picture, any optomechanical interferometer can be characterised by a linear \textit{input-output (I/O) relations} written as a linear vector transformation of the form:
\begin{equation}\label{eq:IOlossless}
  \pmb{b} = \mathbb{T}\pmb{a} + \pmb{t}\frac{h}{h_{\rm SQL}},\mbox{ where }
  \mathbb{T}=\begin{bmatrix}
    T_{cc} & T_{cs}\\
    T_{sc} & T_{ss}
  \end{bmatrix},\ \pmb{t} = \begin{bmatrix}
    t_c\\
    t_s
  \end{bmatrix}\,,
\end{equation}
where $\pmb{a} = [\hat a_c(\Omega),\,\hat a_s(\Omega)]^{\rm T}$ and $\pmb{b} = [\hat b_c(\Omega),\,\hat b_s(\Omega)]^{\rm T}$ are the 2-dimensional vectors of the input and the output light quadratures, respectively, $\mathbb{T}(\Omega)$ is a $2\times2$-matrix of the corresponding optical transfer matrices for the light fields, and $\pmb{t}(\Omega)$ is a 2-dimensional vector of optomechanical response functions that characterises how GW with strain amplitude spectrum $h(\Omega)$ shows itself in the output quadratures of the light leaving the interferometer. 

Readout photocurrent of the balanced homodyne detector is proportional to the quadrature of the outgoing light defined by the local oscillator phase angle $\phi_{LO}$. Thus we can define the readout quadrature proportional to $\hat I_{\rm BHD}(\phi_{LO})$ as :
\begin{equation}\label{eq:o_zeta_lossless}
  \hat{o}_{\phi_{LO}} \equiv \hat{b}_c\cos\phi_{LO} + \hat{b}_s\sin\phi_{LO} \equiv \pmb{\rm H}_{\phi_{LO}}^{\rm T}\cdot\pmb{b}\,,
\end{equation}
with $ \pmb{\rm H}_{\phi_{LO}}\equiv\{\cos\phi_{LO},\,\sin\phi_{LO}\}^{\rm T}$ and the spectral density of quantum noise at the output port of the interferometer can be obtained using the following simple rule:
\begin{equation}\label{eq:SpDens_h}
  S^h(\Omega) = h^2_{\rm SQL}\frac{\pmb{\rm H}^{\rm T}_{\phi_{LO}}\cdot\mathbb{T}\cdot\mathbb{S}_{a}^{in}\cdot\mathbb{T}^\dag\cdot\pmb{\rm H}_{\phi_{LO}}}{|\pmb{\rm H}^{\rm T}_{\phi_{LO}}\cdot\pmb{t}_h|^2}
\end{equation}
where $\mathbb{S}_a^{in}$ stands for spectral density matrix of injected light and components thereof can be defined as:
\begin{multline}\label{eq:SpDens_a}
  2\pi\delta(\Omega-\Omega') \, \mathbb{S}_{a, ij}^{in}(\Omega) \equiv \\
    \frac12\bra{in}\hat{a}_i(\Omega)(\hat{a}_j(\Omega'))^\dag+(\hat{a}_j(\Omega'))^\dag\hat{a}_i(\Omega)\ket{in}\,,
\end{multline}
where $\ket{in}$ is the quantum state of vacuum injected in the dark port of the interferometer and $(i,j) = \{c,s\}$  (see Sec. 3.3 in \cite{Liv.Rv.Rel.15.2012} for more details).
In present article we deal with \emph{single-sided} spectral densities $S$ and hence in case of input vacuum state:
\begin{equation*}
  \ket{in} = \ket{vac} \qquad\Rightarrow\qquad \mathbb{S}_a^{in} = \mathbb{I}\,.
\end{equation*}

\paragraph{I/O-relations of the polarisation circulation speed meter:}
The I/O-relations for our scheme can be obtained, using the Michelson interferometer I/O-relations twice, for each of the $\pm45^\circ$-polarisation modes. One just needs to keep in mind that both polarisations contribute to the common back-action force. The corresponding transfer matrix, $\mathbb{T}_{\rm MI}$ and response vector, $\pmb{t}_{\rm MI}$, read:
\begin{equation}
\mathbb{T}_{\rm MI} = e^{2i\beta_{\rm MI}}\begin{bmatrix}
0 & 1\\
-\mathcal{K}_{\rm MI} & 1
\end{bmatrix},\ 
\pmb{t}_{\rm MI}=e^{i\beta_{\rm MI}}\sqrt{2\mathcal{K}_{\rm MI}}
\begin{bmatrix}
0\\
1
\end{bmatrix}.%
\end{equation}

In the proposed scheme, both polarisation modes, $\pmb{p}_l$  and $\pmb{p}_r$, have half of the total circulating power provided by the pump laser. Therefore, each mode has only half of the full Michelson power and thus $\mathcal{K}_{r,l} \to \mathcal{K}_{\rm MI}/2$. Having this in mind, one can write down the I/O-relations for the two polarisation modes and for the link between them, provided by the PMC unit as:
\begin{align}
\begin{cases}
\pmb{\hat o}_l &= \mathbb{T}^l_{\rm MI}\pmb{\hat i}_l+\mathbb{T}^{\rm b.a.}_{\rm MI}\pmb{\hat i}_r+\pmb{t}_{l}\frac{h}{h_{\rm SQL}}\,,\\
\pmb{\hat o}_r &= \mathbb{T}^{\rm b.a.}_{\rm MI}\pmb{\hat i}_l+\mathbb{T}^r_{\rm MI}\pmb{\hat i}_r+\pmb{t}_{r}\frac{h}{h_{\rm SQL}}\,,\\
\pmb{\hat i}_r &= \mathbb{P}^2_{\phi_0} \pmb{\hat o}_l\,.
\end{cases}%
\end{align}

where $\mathbb{P}_{\phi_0} = \begin{bmatrix}
\cos\phi_0 & -\sin\phi_0\\
\sin\phi_0 & \cos\phi_0
\end{bmatrix}$ is the 2D rotation matrix by angle $\phi_0$ that describes the phase shift carrier light acquires as it propagates from the QWP towards the PCM, and $\mathbb{T}_{\rm MI}^{\rm b.a.} = e^{2i\beta}\begin{bmatrix}
0 & 0\\
-\mathcal{K}/2 & 0
\end{bmatrix}$ is the back-action-only transfer matrix of the arm that accounts for the back-action effect on the corresponding polarisation mode created by the orthogonal mode radiation pressure. 

Solving these equations for $\pmb{\hat o}_r$, one gets for the new transfer matrix, $\mathbb{T}[\phi_0]$, and response function, $\pmb{t}[\phi_0]$:
\begin{align}\label{eq:PCSM_Tmat_tvec_gen}
\mathbb{T}[\phi_0] &= \mathbb{T}_{\rm MI}^{\rm b.a.}+\mathbb{T}^r_{\rm MI}\cdot\mathbb{P}^2_{\phi_0}\cdot\left[\mathbb{I}-\mathbb{T}_{\rm MI}^{\rm b.a.}\cdot\mathbb{P}^2_{\phi_0}\right]^{-1}\cdot\mathbb{T}^l_{\rm MI}\,,\\
\pmb{t}[\phi_0] &= \pmb{t}_{r} + \mathbb{T}^r_{\rm MI}\cdot\mathbb{P}^2_{\phi_0}\cdot\left[\mathbb{I}-\mathbb{T}_{\rm MI}^{\rm b.a.}\cdot\mathbb{P}^2_{\phi_0}\right]^{-1}\cdot\pmb{t}_{l}\,.
\end{align}

The speed meter regime of this interferometer is achieved when $2\phi_0 = \pi n$ for all integer $n$. In this case, one has:
 \begin{equation}\label{eq:PCSM_Tmat}
	  \mathbb{T}=-e^{4i\beta}\begin{bmatrix}
     1 & 0\\
    -2\mathcal{K}\sin^2\beta & 1
  \end{bmatrix} = e^{2i\beta_{\rm sag}}\begin{bmatrix}
  1 & 0\\
  -\mathcal{K}_{\rm sag}/2 & 1
  \end{bmatrix}
\end{equation}
\begin{equation}\label{eq:PCSM_tvec}
	\pmb{t} = e^{2i\beta}\sqrt{\mathcal{K}}\begin{bmatrix}
	   0\\
	    -2i\sin\beta
	  \end{bmatrix} = e^{i\beta_{\rm sag}}\sqrt{\mathcal{K}_{\rm sag}}\begin{bmatrix}
	  0\\
	  1
	  \end{bmatrix}\,,
\end{equation}
where $\mathcal{K}_{\rm sag} = 4\mathcal{K}\sin^2\beta$ is the Sagnac speed meter OM coupling factor and $\beta_{\rm sag} = 2\beta+\pi/2$ is the corresponding phase shift for sidebands travelling through the Sagnac interferometer \cite{Chen2002}. So we have shown that our scheme is equivalent to the Sagnac speed meter interferometer with 2 times lower input laser power. There is no surprise in that. 

And finally, substituting \eqref{eq:PCSM_Tmat} and \eqref{eq:PCSM_tvec} into Eq.~\eqref{eq:SpDens_h}, one gets the final expression for the PCSM quantum noise power spectral density in the form \eqref{eq:Sh_tuned}.

Arbitrary values of $\phi_0$ yield far more cumbersome formulas for $\mathbb{T}$ and $\pmb{t}$ that one can obtain straightforwardly from the Eqs.~\eqref{eq:PCSM_Tmat_tvec_gen}. However, the simple case of small variation of $\phi_0$ from $\pi/2$ value is of special interest for the analysis of the influence of imperfections. Let assume $\phi_0 = \pi/2+\epsilon$ where $\epsilon=2\pi\Delta L_{\rm PC}/\lambda_p\ll1$, then one gets in the first order in $\epsilon$:
\begin{equation}\label{eq:PCSM_Tmat_epsilon}
	  \mathbb{T}_\epsilon=-\frac{e^{4i\beta}}{1+2 e^{2i\beta}\epsilon\mathcal{K}}\begin{bmatrix}
     1+\mathcal{K}\epsilon & -2\epsilon\\
    -2(\mathcal{K}\sin^2\beta-\epsilon) & 1+\mathcal{K}\epsilon
  \end{bmatrix}
\end{equation}
\begin{equation}\label{eq:PCSM_tvec_epsilon}
	\pmb{t}_\epsilon = \frac{2e^{2i\beta}\sqrt{\mathcal{K}}}{1+2 e^{2i\beta}\epsilon\mathcal{K}}\begin{bmatrix}
	   e^{i\beta}\\
	    -i\sin\beta
	  \end{bmatrix}\,,
\end{equation}
that for phase quadrature readout gives the following simple expression for the QNLS PSD:
\begin{equation}\label{eq:PCSM_QNLS_epsilon}
	S^h_\epsilon \simeq \dfrac{h^2_{\rm SQL}}{2}\Bigl\{\dfrac{2}{\mathcal{K}_{\rm sag}}+\dfrac{\mathcal{K}_{\rm sag}}{2}+\dfrac{2\epsilon(\mathcal{K}-\mathcal{K}_{\rm sag})}{\mathcal{K}_{\rm sag}}\Bigr\}\,,
\end{equation}
where the last term in the brackets dominates at low frequencies, being $\propto 1/\Omega^6$, as we discussed above.

\paragraph{Contributions:} N.V., F.Ya. and S.D. conceived the polarisation circulation speed meter concept. S.D. and E.K. carried out the theoretical calculations and numerical modelling of the quantum noise perfromance with contributions from all other authors. S.S., C.G., S.H. and J.H. analysed the experimental feasibility of the assumed imperfections of optical components. S.D. and S.H. performed the astrophycal bench-marking and wrote the manuscript with the contributions of all the authors.

\end{document}